\documentclass[aps,preprint,nofootinbib,superscriptaddress]{revtex4}
\usepackage{amssymb}

\usepackage{epsfig}

\begin{document}

\date{\today}
\title{Applications of Skyrme energy-density functional to fusion reactions for synthesis of superheavy nuclei}

\author{Ning Wang}
\email{Ning.Wang@theo.physik.uni-giessen.de}
\affiliation{Institute for Theoretical Physics at
Justus-Liebig-University, D-35392 Giessen, Germany}

\author{Xizhen Wu}
\affiliation{China Institute of Atomic Energy, Beijing 102413, P.
R. China}

\author{Zhuxia Li}
\affiliation{China Institute of Atomic Energy, Beijing 102413, P.
R. China} \affiliation{Institute of Theoretical Physics, Chinese
Academic of Science, Beijing 100080, P. R. China}
\affiliation{Nuclear Theory Center of National Laboratory of Heavy
Ion Accelerator, Lanzhou 730000, P. R. China}

\author{Min Liu}
\affiliation{China Institute of Atomic Energy, Beijing 102413, P.
R. China}

\author{Werner Scheid}
\affiliation{Institute for Theoretical Physics at
Justus-Liebig-University, D-35392 Giessen, Germany}

\begin{abstract}
The Skyrme energy-density functional approach has been extended to
study the massive heavy-ion fusion reactions. Based on the
potential barrier obtained and the parameterized barrier
distribution the fusion (capture) excitation functions of a lot of
heavy-ion fusion reactions are studied systematically. The average
deviations of fusion cross sections at energies near and above the
barriers from experimental data are less than 0.05 for $92\%$ of
76 fusion reactions with $Z_1Z_2<1200$. For the massive fusion
reactions, for example, the $^{238}$U-induced reactions and
$^{48}$Ca+$^{208}$Pb the capture excitation functions have been
reproduced remarkable well. The influence of structure effects in
the reaction partners on the capture cross sections are studied
with our parameterized barrier distribution. Through comparing the
reactions induced by double-magic nucleus $^{48}$Ca and by
$^{32}$S and $^{35}$Cl, the 'threshold-like' behavior in the
capture excitation function for $^{48}$Ca induced reactions is
explored and an optimal balance between the capture cross section
and the excitation energy of the compound nucleus is studied.
Finally, the fusion reactions with $^{36}$S, $^{37}$Cl, $^{48}$Ca
and $^{50}$Ti bombarding on $^{248}$Cm, $^{247,249}$Bk,
$^{250,252,254}$Cf and $^{252,254}$Es, and as well as the
reactions lead to the same compound nucleus with $Z=120$ and
$N=182$ are studied further. The calculation results for these
reactions are useful for searching for the optimal fusion
configuration and suitable incident energy in the synthesis of
superheavy nuclei.
\end{abstract}

\maketitle


\begin{center}
\textbf{I. INTRODUCTION}
\end{center}

It is of great importance to predict fusion cross sections and to
analyze reaction mechanism for massive heavy-ion fusion reactions,
especially for fusion reactions leading to superheavy nuclei. In
those reactions, the calculation of the capture cross section is
of crucial importance. It is known that Wong's
formula\cite{Wong73} based on one-dimensional barrier penetration
can describe the fusion excitation function well for light
reaction systems, while it fails to give satisfying results for
heavy reaction systems at energies near and below the barrier. For
solving this problem, a fusion coupled channel model \cite{Hag99}
was proposed, in which the macroscopic Woods-Saxon potential
together with a microscopic channel coupling concept is adopted.
By this model fusion excitation functions of some reactions at
energies near and below the barrier are successfully described.
However, it has been found that the parameters in the Woods-Saxon
potential greatly influence on the results \cite{New04} and for
heavy systems the potential parameters need to be readjusted in
order to reproduce experimental data \cite{Siw04}. How to
determine the parameters is still an open problem for predicting
fusion cross sections of unmeasured reaction systems. Therefore,
it is highly requisite to propose a new method for systematically
describing fusion reactions from light to heavy reaction systems.

In our previous paper\cite{Liu05}, we applied the Skyrme
energy-density functional for the first time to study heavy-ion
fusion reactions. The barrier for fusion reaction was calculated
by the Skyrme energy-density functional together with the
semi-classical extended Thomas-Fermi method\cite{brack}. Based on
the interaction potential barrier obtained, we proposed a
parametrization of the empirical barrier distribution to take into
account the multi-dimensional character of the real barrier and
then applied it to calculate the fusion excitation functions of
light and intermediate-heavy fusion reaction systems in terms of
the barrier penetration concept. A large number of measured fusion
excitation functions at energies around the barriers were
reproduced well. Now we try to extend this approach to study very
heavy fusion reaction systems which may lead to the formation of
superheavy nuclei. In these cases, the reaction mechanism is very
complicated and the capture process is the firstly concerned
process, which follows by the quasi-fission and fusion, and then
the fused system further undergoes fusion-fission and evaporation.

The study of the fusion mechanism (or capture process in very
heavy fusion systems), especially of the possible enhancement of
the fusion (capture) cross section in neutron-rich reactions and
also of the suppression of fusion (capture) cross section induced
by the strong shell effects of projectile or target, is very
interesting and essentially importance for the synthesis of
superheavy nuclei. For fusion reactions induced by double-magic
nucleus $^{48}$Ca, there exists a puzzle: on one hand, it has been
found that the fusion cross sections at sub-barrier energies are
suppressed in fusion reactions $^{48}$Ca+$^{48}$Ca\cite{Trot01}
and $^{48}$Ca+$^{90,96}$Zr \cite{Scar04,Zhang05} compared with the
$^{40}$Ca+$^{48}$Ca and $^{40}$Ca+$^{90,96}$Zr, respectively. On
the other hand, the experiments of production of superheavy
elements $Z=114$ and 116 in "hot fusion" reactions with $^{48}$Ca
bombarding Pu and Cm targets\cite{Ogan00} indicate that the
reactions with $^{48}$Ca nuclei, indeed, are quite favorable for
the synthesis of superheavy nuclei. Therefore, it is worthwhile to
explore the puzzle concerning the fusion reactions induced by
$^{48}$Ca. For this purpose, the influence of shell structure,
that is, the influence of the Q-value in fusion (capture) process
on the fusion (capture) cross section is considered in our
approach. The choice of an optimal reaction combination and a
suitable incident energy is always of crucial importance for the
synthesis of new superheavy nuclei. In order to choose a suitable
incident energy, an optimal balance between capture cross section
and excitation energy of compound nuclei should be taken into
account. Thus, in this work, a series of fusion reactions induced
by $^{48}$Ca, $^{36}$S, $^{37}$Cl and $^{50}$Ti are investigated
within our approach and the optimal incident energies for the
reactions are given.

\begin{center}
\textbf{II. MICROSCOPIC INTERACTION POTENTIAL BARRIER AND
PARAMETERIZED BARRIER DISTRIBUTION}
\end{center}

In this section, we briefly introduce our approach for calculating
the interaction potential barrier and fusion (capture) excitation
function, a more detailed description can be found in
ref.\cite{Liu05}. The nucleus-nucleus interaction potentials of
fusion systems are calculated within the microscopic Skyrme
energy-density functional together with the semi-classical
extended Thomas-Fermi (ETF2) approach (up to second order of
$\hbar$). The interaction potential $V_b(R)$ between reaction
partners can be written as
\begin{eqnarray}
V_b(R) = E_{tot}(R) - E_1 - E_2,
\end{eqnarray}
where $R$ is the center-to-center distance between reaction
partners, $E_{tot}(R)$ is the total energy of the interaction
system, $E_1$ and $E_2$ are the energies of the non-interacting
projectile and target, respectively. The interaction potential
$V_b(R)$ is also called the entrance-channel potential in
ref.\cite{Deni02} or fusion potential in ref.\cite{Bass74}. The
$E_{tot}(R)$, $E_1$, $E_2$ are determined by the Skyrme
energy-density functional\cite{Vau72,Bart85,brack,Deni02,Bart02},
\begin{eqnarray}
E_{tot}(R) =  \int \; {\mathcal H}[ \rho_{1p}({\bf
r})+\rho_{2p}({\bf r}-{\bf R}), \rho_{1n}({\bf r})+\rho_{2n}({\bf
r}-{\bf R})] \; d{\bf r},
\end{eqnarray}
\begin{eqnarray}
E_1 = \int \; {\mathcal H}[ \rho_{1p}({\bf r}), \rho_{1n}({\bf
r})] \;
d{\bf r}, \\
E_2 = \int \; {\mathcal H}[ \rho_{2p}({\bf r}), \rho_{2n}({\bf
r})] \; d{\bf r}.
\end{eqnarray}
Here, $\rho_{1p}$, $\rho_{2p}$, $\rho_{1n}$ and $\rho_{2n}$ are
the frozen proton and neutron densities of the projectile and
target, and the expression of the energy-density functional
$\mathcal H$ can be found in refs.\cite{Deni02,Liu05}. Once the
proton and neutron density distributions of the projectile and
target are determined, the interaction potential $V_b(R)$ can be
calculated from eqs.$(1)-(4)$.

By density-variational approach and minimizing the total energy of
a single nucleus given by the Skyrme energy-density functional
$\mathcal H$, the neutron and proton densities of this nucleus can
be obtained. In this work we take the neutron ($i=n$) and proton
($i=p$) density distributions of nuclei as spherical symmetric
Fermi functions,
\begin{equation}
\rho_{i}({\bf r})=\rho _{0i} \left[1+\exp
\left(\frac{r-R_{0i}}{a_{i}}\right)\right]^{-1}, \; \; \;
i=\left\{ {n,p}\right\}.
\end{equation}
Only two of the three quantities $\rho _{0i}$ , $R_{0i}$ and
$a_{i}$ in this relation, are independent because of the
conservation of the particle numbers $N_i = \int \rho_i({\bf r})
d{\bf r} $,   $N_i = \left\{ {N,Z}\right\} $. For example, $\rho
_{0p}$ can be expressed as a function of $R_{0p}$ and $a_{p}$,
\begin{eqnarray}
\rho _{0p} \simeq Z \left\{ \frac{4}{3}\pi R_{0p}^{3}\left[ 1+\pi
^{2}\left( \frac{a_{p}}{R_{0p}}\right)^{2}\right] \right\}^{-1}
\end{eqnarray}
with high accuracy\cite{Gram82} when $R_{0p} \gg a_{p}$. By using
an optimization algorithm, one can obtain the minimal energy and
the corresponding $R_{0p}$, $a_{p}$, $R_{0n}$,$a_{n}$ for neutron
and proton densities. Then, with the neutron and proton densities
of projectile and target obtained we can calculate the
entrance-channel potential with the same energy-density
functional. For systematically investigating massive heavy-ion
fusion reactions with a simple self-consistent manner provided by
the density functional theory\cite{Hoh64}, an optimal balance
between the accuracy and computation cost is adopted in this
approach, which is especially valuable for theses cases.

\begin{figure}
\includegraphics[angle=-0,width=0.8\textwidth]{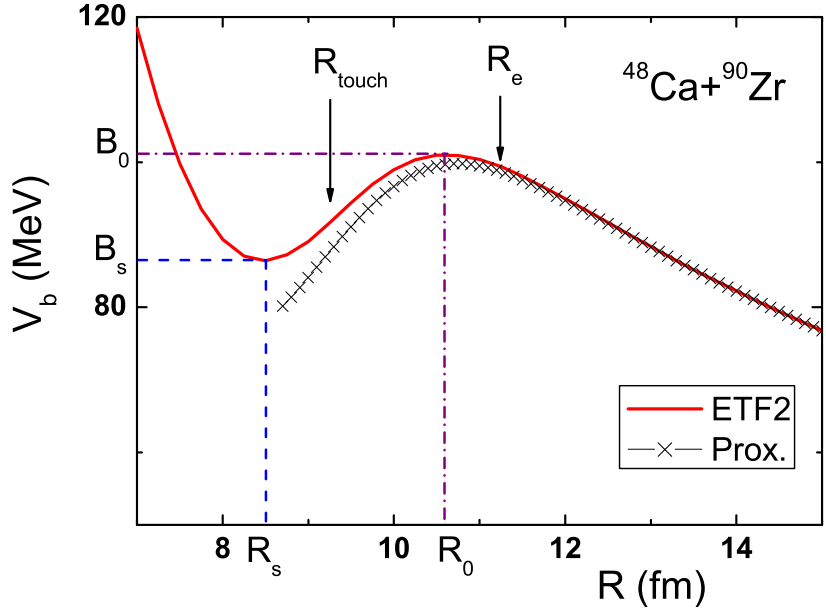}
 \caption{ (Color online) The entrance-channel potential for
reaction $^{48}$Ca+$^{90}$Zr. } \label{fig1}
\end{figure}

The Skyrme force SkM*\cite{Bart85} is adopted in this work. For a
certain reaction system, the entrance-channel potential is
calculated in a range from $R=7fm$ to $15fm$ with a step size
$\Delta R=0.25fm$. Fig.1 shows the entrance-channel potential of
$^{48}$Ca+$^{90}$Zr. The solid and crossed curves denote the
results of this approach and of the proximity
potential\cite{Myers00}, respectively. The results of Skyrme
energy-density functional approach are generally close to those of
proximity potential in the region where the densities of the two
nuclei do not overlap. The barrier height $B_0$, radius $R_0$ and
the curvature $\hbar\omega_0$ near $R_0$ as well as the position
of fusion pocket $R_{\rm s}$ can be obtained from the calculations
(see Fig.1). Here, the curvature $\hbar\omega_0$ of the barrier is
obtained through fitting the entrance-channel potential in the
region from $R_0-1.25fm$ to $R_0+1.25fm$ by an inverted parabola
(if $R_0-1.25fm<R_{\rm s}$, then from $R_{\rm s}$ to $R_0+1.25fm
$).

To overcome the deficiency of one-dimensional barrier penetration
model for describing sub-barrier fusion of heavy systems, we take
into account the multi-dimensional character of realistic
barrier\cite{Stel88} due to the coupling to internal degrees of
freedom of the binary system. We assume that the one-dimensional
barrier is replaced by a distribution of fusion barrier $D(B)$.
The distribution function $D(B)$ satisfies:
\begin{eqnarray}
\int_{0}^{ \infty }D(B)dB=1.
\end{eqnarray}
Motivated by the shape of the barrier distribution extracted from
experiments, we consider the weighting function to be a
superposition of two Gaussian functions $D_1(B)$ and $D_2(B)$,
which read
\begin{eqnarray}
D_{1}(B)=\frac{\sqrt{\gamma}}{2\sqrt{\pi}w_{1}} \exp \left[ -\gamma \frac{%
(B-B_{1})^{2}}{(2w_{1})^{2}}\right]
\end{eqnarray}
and
\begin{eqnarray}
D_{2}(B)=\frac{1}{2\sqrt{\pi}w_{2}}\exp \left[ -\frac{%
(B-B_{2})^{2}}{(2w_{2})^{2}}\right],
\end{eqnarray}
with
\begin{eqnarray}
w_{1}=\frac{1}{4}(B_0-B_{c}),
\end{eqnarray}
\begin{eqnarray}
w_{2}=\frac{1}{2}(B_0-B_{c}),
\end{eqnarray}
\begin{eqnarray}
B_{1}=B_{c}+w_{1},
\end{eqnarray}
\begin{eqnarray}
B_{2}=B_{c}+w_{2}.
\end{eqnarray}
Here $B_0$ is the height of the barrier (see Fig.1). The $B_{c}=f
B_0$ is the effective barrier height with a reducing factor $f$ to
mimic the lowering barrier effect which is due to the coupling to
other degrees of freedom, such as dynamical deformation and
nucleon transfer, etc. We set the reducing factor $f=0.926$ in
this work, which is the same as in \cite{Liu05}. The quantity
$\gamma$ in $D_1(B)$ is a factor to taken into account the
structure effects, which influences the width of the distribution
$D_1(B)$. For the fusion reactions with non-closed-shell nuclei
but near the $\beta$-stability line we set $\gamma=1$; for fusion
reactions with neutron-shell-closed nuclei or neutron-rich nuclei
an empirical formula for the $\gamma$ values, used in the
weighting function $D_1(B)$ for systems with the same $Z_{1}$ and
$Z_{2}$, was proposed in ref.\cite{Liu05} as
\begin{eqnarray}
\gamma=1-c_0 \Delta Q + 0.5(\delta_{n}^{prog}+\delta_{n}^{targ}),
\end{eqnarray}
where $\Delta Q=Q-Q_0$ denotes the difference between the Q-values
of the system under considering for complete fusion and that of
the reference system. The reference system, in general, is chosen
to be the reaction system with nuclei along the $\beta$-stability
line \cite{Liu05}. The value of $c_0$ is $0.5 MeV^{-1}$ for
$\Delta Q<0$ and $0.1MeV^{-1}$ for $\Delta Q>0$. The quantities
$\delta_{n}^{proj(targ)}$ are 1 for neutron closed-shell
projectile (target) nucleus and 0 for non-closed cases.

The fusion excitation function is then given by
\begin{eqnarray}
\sigma_f (E_{\rm c.m.})=\int_{0}^{ \infty }D(B) \sigma _{fus}^{\rm
Wong}(E_{\rm c.m.},B)dB,
\end{eqnarray}
with
\begin{equation}
\sigma _{ fus}^{\rm Wong}(E_{\rm c.m.},B)=\frac{\hbar \omega
_{0}R_{0}^{2}}{2E_{\rm c.m.}}\ln \left( 1+\exp\left[ \frac{2\pi
}{\hbar \omega _{0}}(E_{\rm c.m.}-B)\right] \right),
\end{equation}
where $E_{\rm c.m.}$ denotes the center-of-mass energy, and $B$,
$R_0$ and $\hbar\omega_0$ are the barrier height, radius and
curvature, respectively. Using the parameterized barrier
distribution functions $D_1(B)$ and $D_2(B)$, we also can obtain
the cross sections $\sigma_1(E_{\rm c.m.})$ and $\sigma_{\rm
avr}(E_{\rm c.m.})$ by (15) with $D(B)$ taken to be $D_1(B)$ and
$D_{\rm avr}(B)=[D_1(B)+D_2(B)]/2$, respectively. Finally, the
fusion cross section is given by
\begin{eqnarray}
\sigma_{fus}(E_{\rm c.m.})=\min[\sigma_1(E_{\rm c.m.}),
\sigma_{\rm avr}(E_{\rm c.m.})].
\end{eqnarray}
The cross section calculated with (17) is referred to as fusion
cross section for a light and intermediate-heavy system and as
capture cross section for a very heavy system at $E_{c.m.}$.

\begin{center}
\textbf{III.  CALCULATED RESULTS FOR FUSION (CAPTURE) EXCITATION
FUNCTIONS}
\end{center}

In order to extend our approach to study the fusion reactions
leading to superheavy nuclei, we first check the suitability and
reliability of our description of heavy-ion fusion reactions. We
calculate the fusion excitation functions of 76 fusion reactions
with $Z_1Z_2 < 1200$ at energies near and above the barrier (with
$\gamma=1$) and their average deviations $\allowbreak \chi _{\log
}^{2}$ from experimental data defined as
\begin{eqnarray}
\allowbreak \chi _{\log }^{2}=\frac{1}{m}\sum_{n=1}^{m} \left[
\log (\sigma_{th}(E_n)) -\log (\sigma_{exp}(E_n))\right]^2.
\end{eqnarray}
Here $m$ denotes the number of energy-points of experimental data,
and $\sigma_{th}(E_n)$ and $\sigma_{exp}(E_n)$ are the calculated
and experimental fusion cross sections at the center-of-mass
energy $E_n$ ($E_n \ge B_{0}$), respectively. Fig.2 shows the
results for  $\allowbreak \chi _{\log }^{2}$ in which the solid
circles and crosses denote the calculated results from this
approach and those from ref.\cite{Hag99}, respectively. Applying
the approach of ref.\cite{Hag99} there are $43 \%$ systems in 76
fusion reactions in which the average deviations $\allowbreak \chi
_{\log }^{2}$ of calculated fusion cross sections from the
experimental data are less than 0.05, but for reactions with
$Z_1Z_2 > 640$ the results are not satisfying very well.
\begin{figure}
\includegraphics[angle=-0,width=1.0\textwidth]{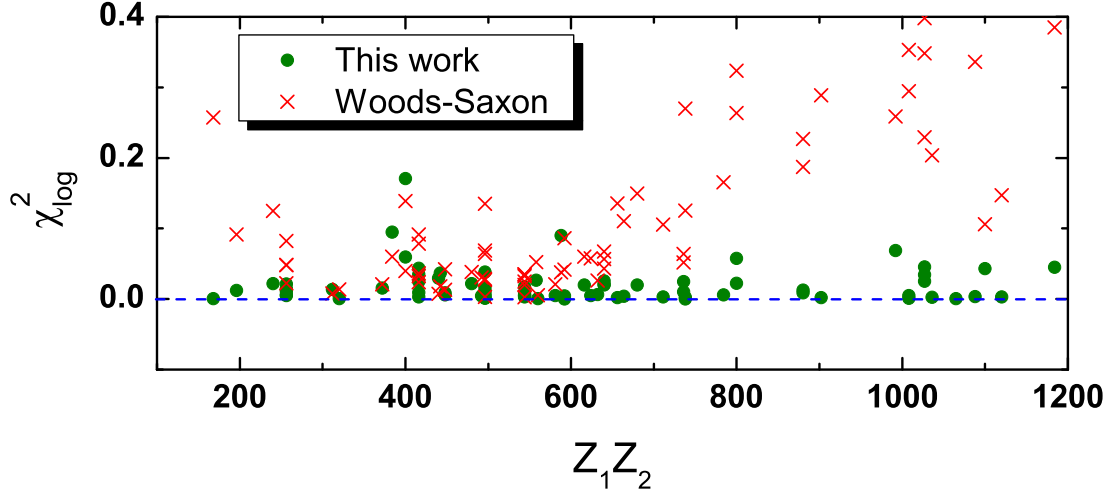}
 \caption{ (Color online) The average deviations $\allowbreak
\chi _{\log }^{2}$ for a total of 76 fusion reactions with $Z_1Z_2
< 1200$. The solid circles and crosses denote the results of our
approach and those with a Woods-Saxon potential with fixed
potential parameters\cite{Hag99}, respectively. In the
calculations of fusion cross sections at energies near and above
the barrier with the Woods-Saxon potential, the code
CCFULL\cite{Hag99} is used without taking into account the
excitation and deformation of the reaction partners.} \label{fig2}
\end{figure}
With our approach, the average deviations of $92 \%$ systems in
$\allowbreak \chi _{\log }^{2}$ are less than 0.05, which
indicates that this approach is successful for describing fusion
cross sections of heavy-ion reactions at energies near and above
the barrier from light to intermediate-heavy fusion systems. In
Fig.3 we show three examples of fusion excitation functions for
the reactions $^{16}$O+$^{144}$Sm\cite{13},
$^{16}$O+$^{92}$Zr\cite{3} and $^{64}$Ni+$^{92}$Zr\cite{Wolf89},
in which the solid and dashed curves present the results of our
approach and of ref.\cite{Hag99}, respectively, the squares denote
the experimental data. From this figure we can see that our
approach gives quite reasonable description for all selected
fusion reactions with the $Z_1Z_2$ up to 1120 at energies near and
above the barrier.

\begin{figure}
\includegraphics[angle=-0,width=1.0\textwidth]{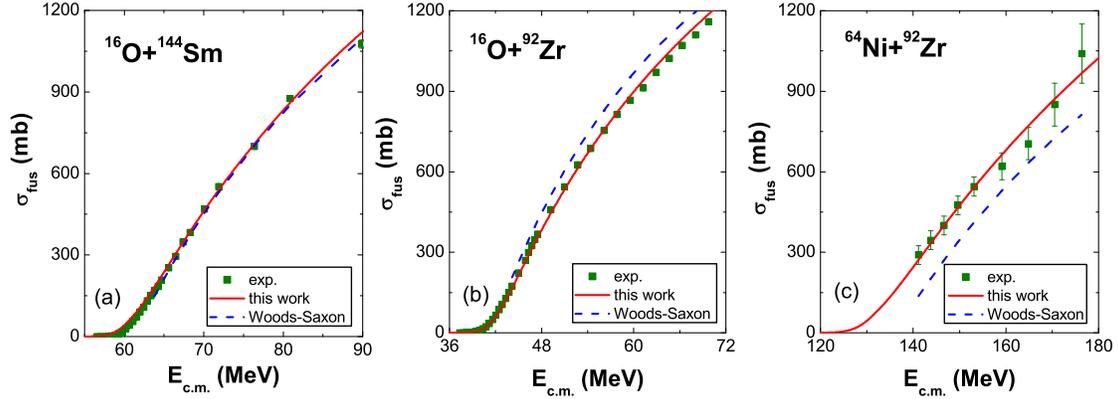}
 \caption{ (Color online) Fusion excitation functions for
$^{16}$O+$^{144}$Sm, $^{16}$O+$^{92}$Zr and $^{64}$Ni+$^{92}$Zr.
The squares and solid curves denote the experimental data and the
results of this work, respectively. The dashed curves denote the
results of the approach with the Woods-Saxon potential with fixed
potential parameters\cite{Hag99}. } \label{fig3}
\end{figure}

For more massive fusion reactions leading to superheavy nuclei,
the quasi-fission process occurs and therefore, the capture cross
sections are larger than the corresponding fusion cross sections.
In ref.\cite{Shen87} the fission and quasi-fission process in
$^{238}$U-induced reactions were studied. Fig.4 shows the results
in which the solid and open circles denote the measured cross
sections for the fission-like process and for complete fusion
followed by fission, respectively. The solid curves give the
calculated results of our approach with $\gamma=1$. From this
figure one can see that the calculated capture excitation
functions of the reactions $^{238}$U+$^{26}$Mg,
$^{238}$U+$^{27}$Al, $^{238}$U+$^{32}$S, and $^{238}$U+$^{35}$Cl
are quite close to the measured fission-like cross sections. It
implies that our approach can describe the massive fusion
reactions between nuclei with neutron open shells but near the
$\beta$-stability line.

\begin{figure}
\includegraphics[angle=-0,width=0.9\textwidth]{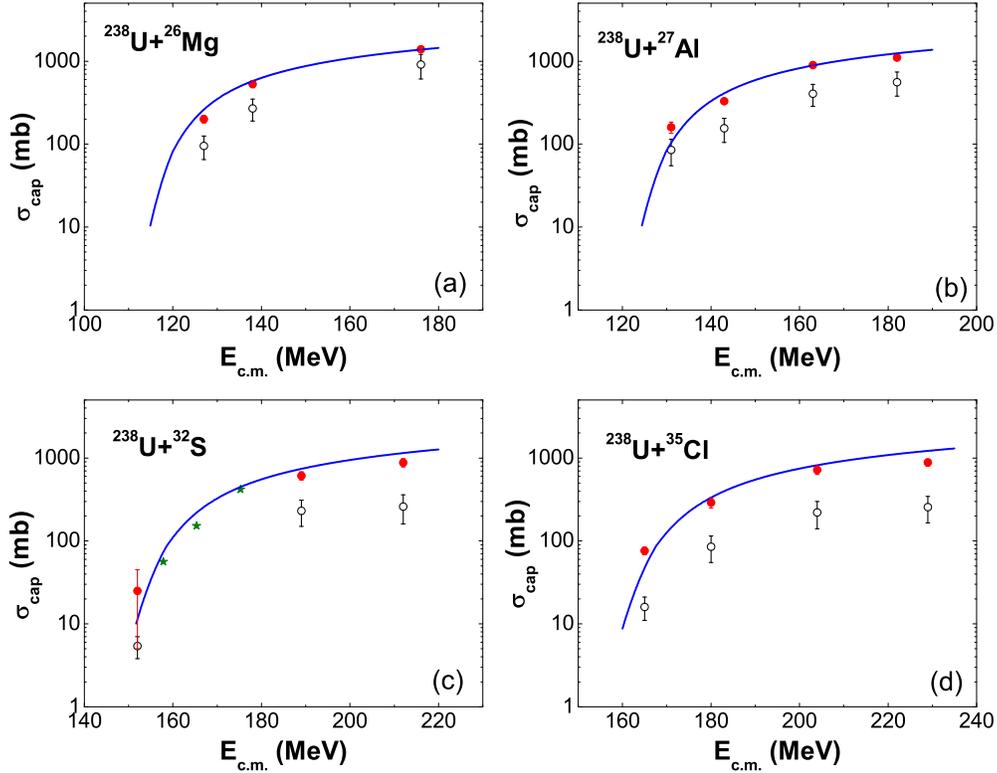}
 \caption{ (Color online) Capture cross sections of
$^{238}$U+$^{26}$Mg,$^{27}$Al,$^{32}$S,$^{35}$Cl. The solid and
open circles denote the measured cross sections for a fission-like
process and for complete fusion followed by fission, respectively.
The solid curves are the results from our approach with
$\gamma=1$. The stars are taken from ref.\cite{Toke84}.}
\label{fig4}
\end{figure}

For the very massive fusion reaction between double-magic nuclei
$^{48}$Ca and $^{208}$Pb, the influence of the shell effects is
very significant. So careful consideration of the $\gamma$ value
is required at sub-barrier energies. Fig.5 shows the calculated
capture excitation function of $^{48}$Ca+$^{208}$Pb and the
experimental data of refs.\cite{Prok03} and \cite{Pach92}. The
dashed curve presents the results with $\gamma=1$, that is, no
neutron-shell-closure effect is considered. The solid curve is
calculated with $\gamma=9.5$ according to eq.(14), in which the
closed shell effect is considered. We find that for energies below
the barrier the experimental data can only be described with
$\gamma=9.5$ and the calculations with $\gamma=1$ are
over-predicted. From this analysis, one learns that the measured
capture cross sections of $^{48}$Ca+$^{208}$Pb at sub-barrier
energies are obviously suppressed, which may arise from the
suppression of the nucleon transfer between reaction partners due
to the strong closed shell effects, which will be further studied
in the following section.

\begin{figure}
\includegraphics[angle=-0,width=0.8\textwidth]{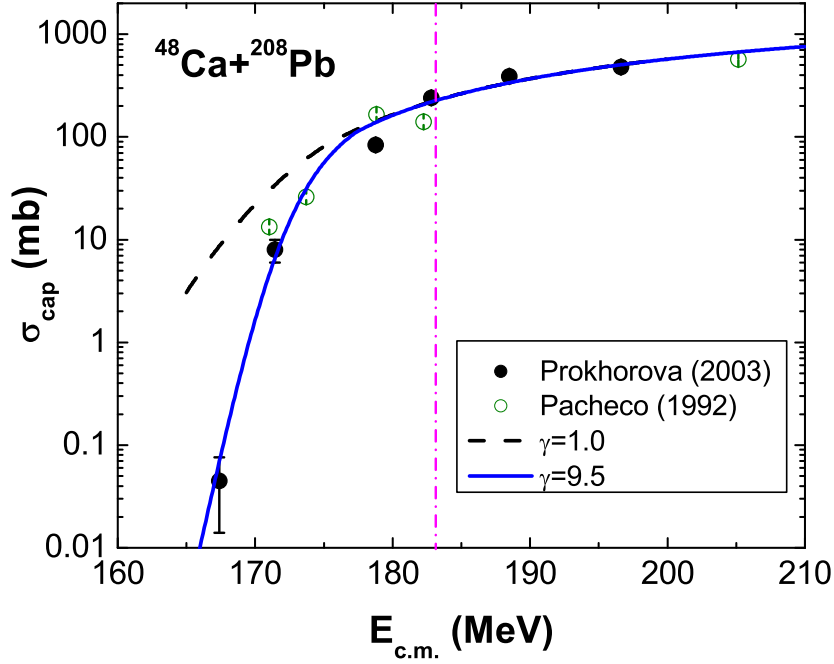}
 \caption{ (Color online) Capture cross sections of
$^{48}$Ca+$^{208}$Pb. The solid and open circles denote the
measured capture-fission cross sections from ref.\cite{Prok03} and
ref.\cite{Pach92}, respectively. The dashed and solid curves
present the calculated results with $\gamma=1.0$ and 9.5 obtained
by eq.(14), respectively. The dash-dotted line indicates the
energy corresponding to the height of the barrier.} \label{fig5}
\end{figure}

\begin{center}
\textbf{IV. OPTIMAL BALANCE BETWEEN CAPTURE CROSS SECTION AND
EXCITATION ENERGY OF COMPOUND NUCLEI}
\end{center}

It is very important to find a favorable combination of projectile
and target and a suitable incident energy for synthesis of
superheavy nuclei. In this section we study very massive fusion
reactions and search for an optimal balance between the capture
cross section in the entrance channel and the excitation energy of
the compound nuclei. For searching a fusion system with large
capture cross sections, we carried out a series of calculations
for fusion reactions induced by $^{32,36}$S, $^{35,37}$Cl and
$^{48}$Ca projectiles. For example, Fig.6 shows the capture
excitation functions of the reactions $^{32}$S+$^{254}$Cf and
$^{35}$Cl+$^{254}$Es. The solid curves present the results with
$\gamma=1$ (without considering structure effects in the entrance
channel), and the dashed curves are for the results with the
$\gamma$ obtained from (14) i.e. $\gamma=0.5$ for
$^{32}$S+$^{254}$Cf and $\gamma=0.6$ for $^{35}$Cl+$^{254}$Es. The
enhancement of capture cross sections in the sub-barrier energy
region with the $\gamma<1$ is caused by the effect of excess of
neutrons in reaction systems. So from the point of view of
increasing the capture cross sections, it is more favorable to
select the reaction systems with $\gamma<1$.
\begin{figure}
\includegraphics[angle=-0,width=1.0\textwidth]{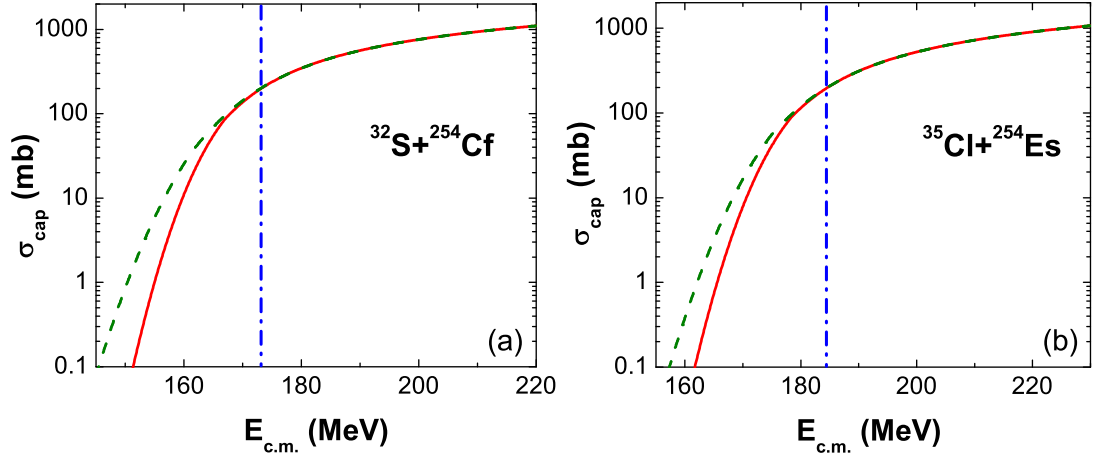}
\caption{ (Color online) Capture excitation functions for fusion
systems $^{32}$S+$^{254}$Cf and $^{35}$Cl+$^{254}$Es.
 The dash-dotted lines indicate the corresponding barriers.
 The solid and dashed curves denote the results with $\gamma=1$ and
 with $\gamma$ value obtained with eq.(14), respectively. }
\end{figure}
However, the amount of the excitation energy of the formed
compound nucleus is essentially important for the survival
probability. The smaller the excitation energy is, the larger the
surviving probability is. Thus, seeking an optimal balance between
the capture cross section and the excitation energy of the
compound nucleus becomes very important for synthesis of
superheavy nuclei . For choosing the fused nuclei with an
excitation energy as low as possible, the fusion reactions with
double-magic nuclei $^{48}$Ca are considered to be good candidates
because of the low Q-values for those fusion reactions. As an
example, let us investigate reaction $^{48}$Ca+$^{248}$Cm. For
this reaction the $\gamma$ value is equal to 10.8 calculated with
eq.(14). Fig.7 shows the capture excitation function for this
reaction, in which the solid and dashed curves denote the results
for the cases of $\gamma$ = 1 and $\gamma$=10.8, respectively.
\begin{figure}
\includegraphics[angle=-0,width=1.0\textwidth]{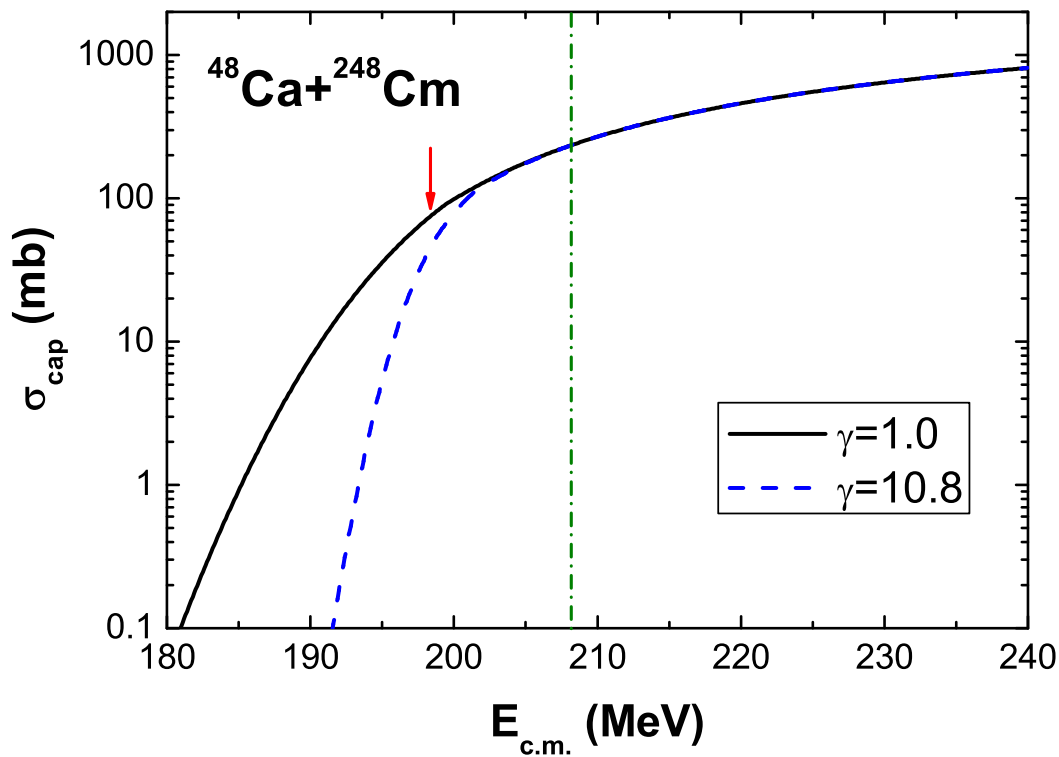}
\caption{ (Color online) Capture excitation functions for
$^{48}$Ca+$^{248}$Cm. The solid and dashed curves represent the
results with the $\gamma$ =1 and with $\gamma$ =10.8 obtained from
eq.(14). The arrow indicates the incident energy at which the
corresponding excitation energy of the formed compound nucleus is
$E^{*}_{CN}=31MeV$. }
\end{figure}
From this figure one finds that for fusion reactions induced by
double-magic nuclei $^{48}$Ca the capture cross sections at
sub-barrier energies are suppressed compared with reactions with
shell-open nuclei but near the $\beta$-stability line. However, if
we suitably choose an incident energy, for example, as indicated
by the arrow in Fig.7, the capture cross section of the reaction
$^{48}$Ca+$^{248}$Cm is not suppressed so much (still reaches
several tens of milli-barns) and the excitation energy of the
compound nuclei is only $E^{*}_{CN}=31MeV$. Such an incident
energy was already used in the experiment of ref.\cite{Ogan00}.
Now let us make a comparison between the reaction
$^{48}$Ca+$^{248}$Cm and the reactions $^{32}$S+$^{254}$Cf and
$^{35}$Cl+$^{254}$Es. For the system $^{48}$Ca+$^{248}$Cm, the
capture cross section is of about $80 mb$ and the excitation
energy is about $31 MeV$ if the incident energy is taken to be
about $198 MeV$. While, for the systems $^{32}$S+$^{254}$Cf and
$^{35}$Cl+$^{254}$Es, if the same excitation energy is required
the incident energies must be as low as about $150 MeV$ and $160
MeV$, respectively, since the Q-values of these two fusion
reactions are much higher compared with $^{48}$Ca induced
reactions. At these incident energies the capture cross sections
for these two reactions are as small as those less than $0.1 mb$
according to this model calculations. From above analysis we can
conclude that the fusion reaction $^{48}$Ca+$^{248}$Cm seems to be
more favorable compared to $^{32}$S+$^{254}$Cf and
$^{35}$Cl+$^{254}$Es if a suitable incident energy is chosen, as
far as both the capture cross section and the excitation energy of
the compound nuclei are concerned.

Now let us discuss how to choose suitable incident energy. We
notice that the capture excitation function for reactions induced
by double-magic $^{48}$Ca goes very sharply down at sub-barrier
energies due to strong closed shell effects, as shown by the
dashed curve of Fig.7. It seems to us that there exists a
'threshold-like' behavior, which is important for choosing the
incident energies. This 'threshold-like' behavior of excitation
function of capture cross sections is closely related to the shape
of the barrier distribution. In our previous paper \cite{Liu05}, a
number of barrier distributions were calculated according to
expressions (8)-(13). As example, here we show the calculated
fusion barrier distribution for $^{16}$O+$^{208}$Pb \cite{Mor99}
in Fig.8. The agreement of the calculated barrier distribution
with experimental data tells us that our approach about the
parameterized barrier distribution is quite reasonable.
\begin{figure}
\includegraphics[angle=-0,width=0.8\textwidth]{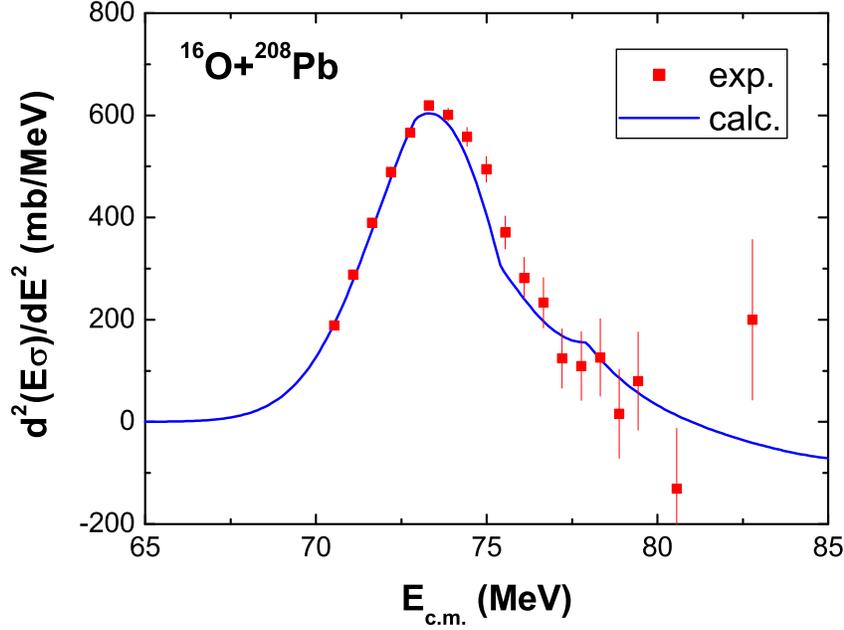}
\caption{ (Color online) Fusion barrier distribution for
$^{16}$O+$^{208}$Pb. The distribution is evaluated with $\Delta
E_{\rm c.m.}=2.5MeV$. The solid squares and solid curve present
the experimental data and the results from our calculations,
respectively.}
\end{figure}
The effective weighting function $D_{\rm eff} (B)$ is defined as
\begin{eqnarray}
D_{\rm eff}(B)= \left\{
\begin{array} {r@{\quad:\quad}l}
D_1(B)& B<B_x \\
D_{\rm avr}(B)& B \ge B_x
\end{array} \right.
\end{eqnarray}
(with  $\int D_{\rm eff}(B)\; dB \approx 1$ and $\int D_{\rm
avr}(B)\; dB = 1$, see ref.\cite{Liu05}). The $B_{x}$ denotes the
position of the left crossing point between $D_1(B)$ and $D_{\rm
avr}(B)$. The function $D_{\rm eff} (B)$ can describe the fusion
excitation function reasonably well. Fig.9 shows the capture
excitation function (Fig.9(a)) and the effective weighting
function $D_{\rm eff}$ (Fig.9(b)) for the reaction
$^{48}$Ca+$^{244}$Pu. The dotted vertical line denotes the barrier
height $B_0$, and the short dashed vertical line indicates the
energy at the peak of $D_{\rm eff}$ which we call the most
probable barrier height $B_{\rm m.p.}$. From the dashed curve of
Fig.9(a) one can see that the capture cross section goes down very
sharply when the incident energy is lower than $B_{\rm m.p.}$.
This is because the decreasing slope of the left side of the
weighting function $D_{\rm eff}$ is very steep due to strong
closed shell effects ($\gamma=11.0$). In fact, one can find that
the left side of the barrier distribution $D_{\rm eff}(B)$ is
given by $D_{1}(B)$ (see expression (19)), which becomes a
$\delta$-function when $\gamma\rightarrow\infty$. For the system
with $\gamma$ much larger than 1 the effective barrier $D_{\rm
eff}$ will have the similar behavior as is shown in Fig.9(b).
Thus, the most probable barrier energy $B_{\rm m.p.}$ can be
considered as the incident energy 'threshold', and for massive
fusion reactions with $\gamma$ much larger than 1 leading to
superheavy nuclei such as $^{48}$Ca induced reactions, the
suitable incident energy should be chosen in the region $E_{\rm
c.m.}>B_{\rm m.p.}$. The barrier distribution for this case shown
in Fig.9(b) looks like a $\delta$-function with a long tail in the
high energy side. It seems to be that the Wong's formula with
barrier height being the $B_{\rm m.p.}$ should work without
introducing the $\gamma$. But the results calculated with Wong's
formula and expression (17) are different especially at
sub-barrier energies, as shown in Fig.9(a) (comparing the
dot-dashed curve and the dashed curve). It seems to us that with
this $\gamma$ value like $\gamma=11.0$ the behavior of $D_{\rm
eff}$ is still different from a $\delta$-function and the
parameter $\gamma$ still plays a role.

\begin{figure}
\includegraphics[angle=-0,width=1.0\textwidth]{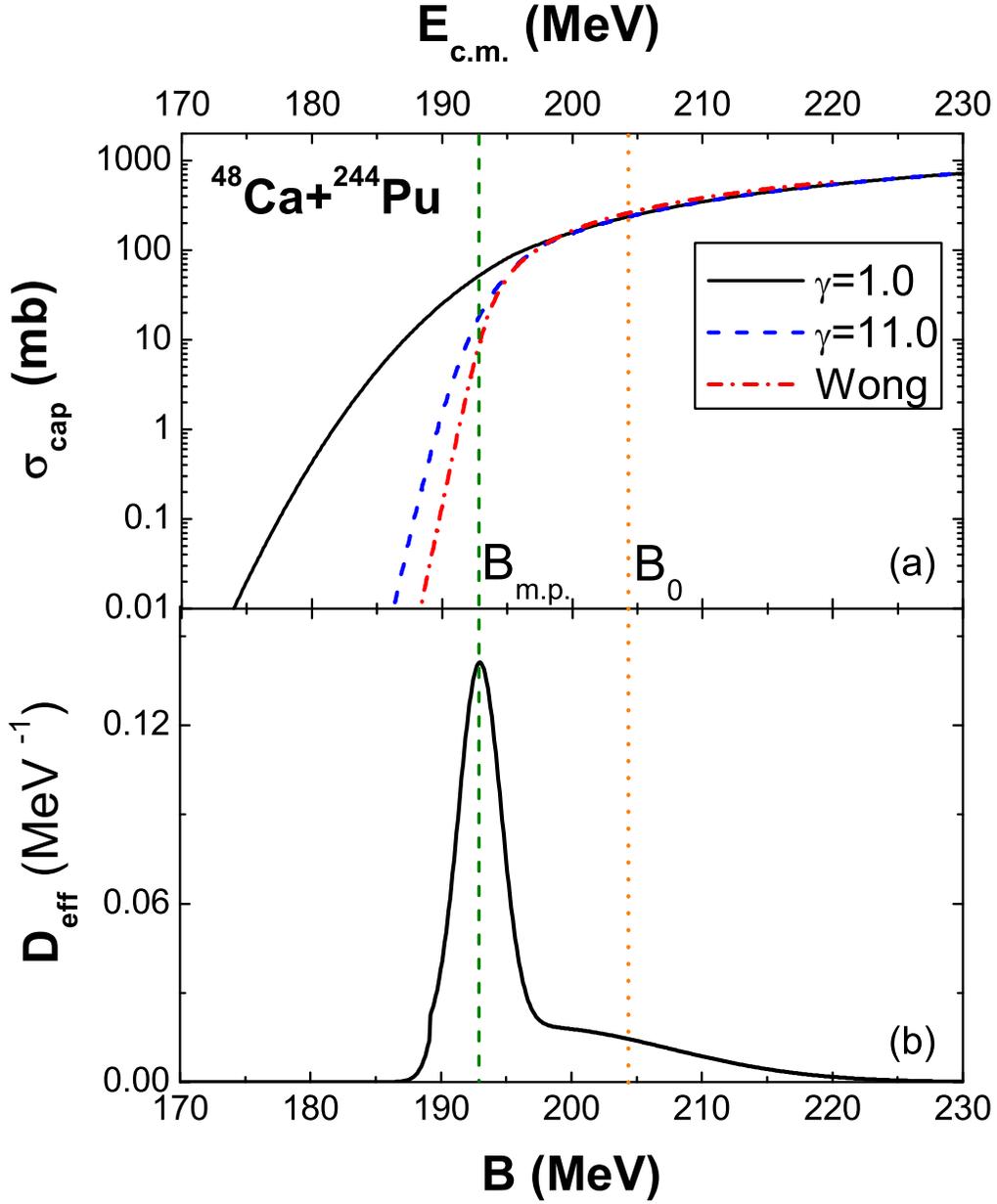}
\caption{ (Color online) (a) Capture excitation function and (b)
effective weighting function for the reaction
$^{48}$Ca+$^{244}$Pu. In (a) the solid and dashed curves show the
results with $\gamma=1$ and with $\gamma$ obtained by eq.(14),
respectively. The dot-dashed curve denotes the results from Wong's
formula with $B=B_{\rm m.p.}$. The results in (b) are obtained by
setting $\gamma=11.0$. }
\end{figure}
We find that the incident energies adopted in the experiments
successfully producing superheavy nuclei in recent years
\cite{CaPb,U238,Pu244,Am243} for some reactions induced by
$^{48}$Ca are very close the most probable barrier energies
$B_{\rm m.p.}$. Table 1 gives the comparison of the calculated
most probable barrier energies $B_{\rm m.p.}$ with experimental
incident energies $E^{exp}_{min}$ used in recent years
\cite{CaPb,U238,Pu244,Am243} for some reactions induced by
$^{48}$Ca leading to producing superheavy nuclei. The barrier
hight $B_0$, position $R_0$ of the barrier, curvature at the top
of the barrier expressed by $\hbar \omega_0$, factor $\gamma$ are
also listed. In addition, we list the mean value $B_{\rm mean}$ of
the barrier height defined as
\begin{eqnarray}
B_{\rm mean}=\frac{\int B \; D_{\rm eff}(B)\; dB}{\int D_{\rm
eff}(B)\; dB}.
\end{eqnarray}
The $B_{\rm mean}$ is, in general, larger than the $B_{\rm m.p.}$
since the slope of the left side of the weighting function $D_{\rm
eff}$ is very steep. From the table one can find that for all
listed reactions the energies $E^{exp}_{min}$ are higher than the
calculated most probable barrier energies $B_{\rm m.p.}$, which
supports our ideas about how to choose the favorable incident
energy. Further, we find the experimental evaporation-residue
excitation functions of the fusion reactions listed in Table 1 are
peaked at the energies ranging from $B_{\rm mean}$ to $B_0$ in
most cases, which implies that the energy $B_{\rm mean}$ may be
more suitable to be chosen as the incident beam energy in the
fusion reactions with $\gamma$ much larger than 1 for producing
superheavy nuclei.

\begin{figure}
\includegraphics[angle=-0,width=1.0\textwidth]{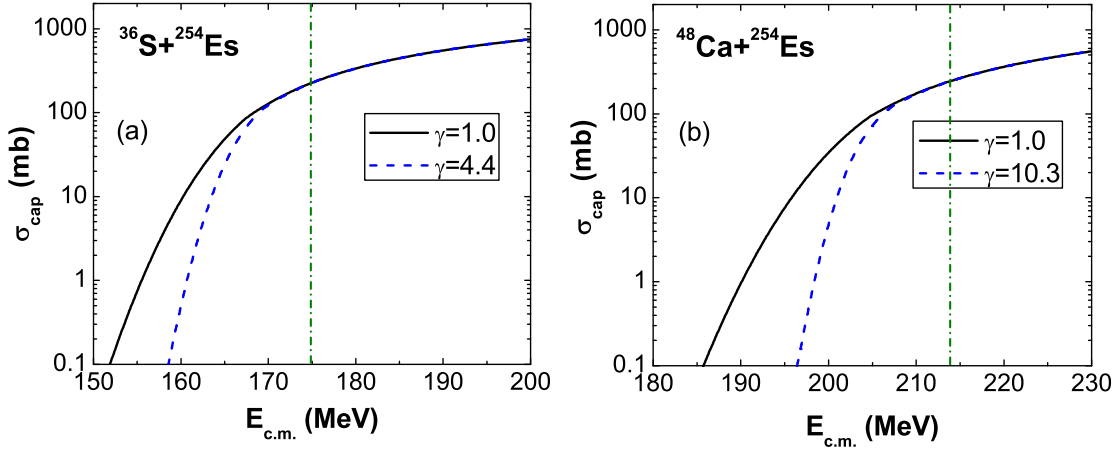}
\caption{ (Color online) Capture excitation functions for the
systems (a) $^{36}$S+$^{254}$Es and (b) $^{48}$Ca+$^{254}$Es.}
\end{figure}

\begin{table}
\caption{ The entrance-channel capture barriers of fusion
reactions with $^{48}$Ca nuclei.}
\begin{tabular}{cccccccc}
\hline\hline
 Reaction &  $ B_0(MeV)$ & $R_0(fm)$ & $\hbar \omega_0(MeV)$ & $\gamma$ & $B_{\rm mean}(MeV)$  & $B_{\rm m.p.}(MeV)$  & $E^{exp}_{min}(MeV)$ \\ \hline
$^{48}$Ca+$^{207}$Pb \cite{CaPb}  & 183.37 & 12.0  & 4.44 & 8.9  & 176.28 & 173.18 & 173.3 \\
$^{48}$Ca+$^{208}$Pb \cite{CaPb}  & 183.17 & 12.0  & 4.43 & 9.5  & 176.10 & 173.03 & 173.5 \\
$^{48}$Ca+$^{238}$U  \cite{U238}  & 200.82 & 12.25 & 4.19 & 10.7 & 193.09 & 189.71 & 191.1 \\
$^{48}$Ca+$^{242}$Pu \cite{U238} & 204.78 & 12.25 & 3.90 & 11.6 & 196.91 & 193.44 & 196   \\
$^{48}$Ca+$^{244}$Pu \cite{Pu244}  & 204.31 & 12.25 & 3.99 & 11.0 & 196.44 & 192.99 & 193.3 \\
$^{48}$Ca+$^{243}$Am \cite{Am243} & 206.87 & 12.25 & 3.87 & 9.8  & 198.89 & 195.39 & 207.1 \\
$^{48}$Ca+$^{245}$Cm \cite{Pu244} & 208.80 & 12.25 & 3.88 & 11.7 & 200.77 & 197.22 & 203   \\
$^{48}$Ca+$^{248}$Cm \cite{U238}  & 208.25 & 12.25 & 3.89 & 10.8 & 200.23 & 196.71 & 198.6 \\
\hline\hline
\end{tabular}
\end{table}

\begin{table}
\caption{ The entrance-channel capture barriers for fusion
reactions with $^{36}$S, $^{37}$Cl, $^{48}$Ca and $^{50}$Ti
bombarding on $^{248}$Cm, $^{247,249}$Bk, $^{250,252,254}$Cf and
$^{252,254}$Es.}

\begin{tabular}{cccccccccc}
\hline\hline
 Reaction & Q$(MeV)$ & $ B_0(MeV)$ & $R_0(fm)$ & $\hbar \omega_0(MeV)$ & $\gamma$ & $B_{\rm mean}(MeV)$  & $B_{\rm m.p.}(MeV)$ & $E^{\ast}_{CN}(MeV)$ & $B_0$-$B_s$  \\ \hline

$^{36}$S+$^{248}$Cm     &   -122.05     &   170.45  &   12.0   &   4.34  & 5.3 & 163.75 & 161.00  &   41.70   &   8.72    \\
$^{36}$S+$^{247}$Bk     &   -126.30     &   172.60  &   12.0   &   4.33  & 4.7 & 165.78 & 163.10  &   39.48   &   8.43    \\
$^{36}$S+$^{249}$Bk     &   -124.58     &   172.03  &   12.0   &   4.23  & 3.9 & 165.18 & 162.58  &   40.60    &   8.23    \\
$^{36}$S+$^{250}$Cf     &   -127.14     &   174.04  &   12.0   &   4.29  & 6.2 & 167.24 & 164.37  &   40.10   &   8.27    \\
$^{36}$S+$^{252}$Cf     &   -125.00     &   173.53  &   12.0   &   4.23  & 5.1 & 166.70 & 163.89  &   41.70   &   8.41    \\
$^{36}$S+$^{254}$Cf     &   -122.48     &   173.02  &   12.0   &   4.17  & 3.9 & 166.13 & 163.51  &   43.65   &   8.49    \\
$^{36}$S+$^{252}$Es     &   -128.84     &   175.36  &   12.0   &   4.23  & 5.4 & 168.47 & 165.65  &   39.63   &   8.18    \\
$^{36}$S+$^{254}$Es     &   -126.81     &   174.97  &   12.0   &   4.17  & 4.4 & 168.04 & 165.26  &   41.23   &   8.25    \\
\hline
$^{37}$Cl+$^{248}$Cm    &   -128.14     &   180.65  &   12.25   &   4.58 & 3.0 & 173.37 & 170.66  &   45.23   &   7.86    \\
$^{37}$Cl+$^{247}$Bk    &   -131.56     &   182.90  &   12.0    &   4.25 & 2.8 & 175.50 & 172.83  &   43.94   &   7.46    \\
$^{37}$Cl+$^{249}$Bk    &   -129.56     &   182.60  &   12.25   &   4.54 & 1.8 & 175.03 & 172.61  &   45.47   &   7.59    \\
$^{37}$Cl+$^{250}$Cf    &   -134.39     &   184.45  &   12.25   &   4.55 & 4.1 & 177.12 & 174.28  &   42.73   &   7.35    \\
$^{37}$Cl+$^{252}$Cf    &   -131.94     &   184.03  &   12.25   &   4.54 & 2.9 & 176.60 & 173.87  &   44.65   &   7.56    \\
$^{37}$Cl+$^{254}$Cf    &   -129.40     &   183.57  &   12.25   &   4.52 & 1.6 & 175.91 & 173.53  &   46.52   &   7.74    \\
$^{37}$Cl+$^{252}$Es    &   -135.20     &   185.96  &   12.25   &   4.54 & 3.5 & 178.51 & 175.70  &   43.31   &   7.23    \\
$^{37}$Cl+$^{254}$Es    &   -132.96     &   185.62  &   12.25   &   4.52 & 2.4 & 178.04 & 175.44  &   45.08   &   7.41    \\
\hline
$^{48}$Ca+$^{248}$Cm    &   -167.27     &   208.25  &   12.25   &   3.89 & 10.8& 200.23 & 196.71  &   32.96   &   5.46    \\
$^{48}$Ca+$^{247}$Bk    &   -171.71     &   210.80  &   12.25   &   3.95 & 9.6 & 202.67 & 199.11  &   30.95   &   5.27    \\
$^{48}$Ca+$^{249}$Bk    &   -170.76     &   210.46  &   12.25   &   3.83 & 9.1 & 202.33 & 198.76  &   31.57   &   5.30    \\
$^{48}$Ca+$^{250}$Cf    &   -174.53     &   212.56  &   12.25   &   3.66 & 11.4& 204.39 & 200.81  &   29.86   &   5.11    \\
$^{48}$Ca+$^{252}$Cf    &   -173.77     &   212.10  &   12.5    &   4.41 & 11.0& 203.94 & 200.37  &   30.17   &   5.17    \\
$^{48}$Ca+$^{254}$Cf    &   -173.28     &   211.63  &   12.5    &   4.38 & 10.7& 203.48 & 199.92  &   30.20   &   5.25    \\
$^{48}$Ca+$^{252}$Es    &   -177.43     &   214.29  &   12.5    &   4.43 & 10.6& 206.04 & 202.39  &   28.61   &   4.98    \\
$^{48}$Ca+$^{254}$Es    &   -176.97     &   213.94  &   12.5    &   4.39 & 10.3& 205.70 & 202.13  &   28.73   &   5.05    \\

\end{tabular}
\end{table}
\begin{table}
\begin{tabular}{cccccccccc}
\hline
  Reaction & Q$(MeV)$ & $ B_0(MeV)$ & $R_0(fm)$ & $\hbar \omega_0(MeV)$ & $\gamma$ & $B_{\rm mean}(MeV)$  & $B_{\rm m.p.}(MeV)$ & $E^{\ast}_{CN}(MeV)$ & $B_0$-$B_s$  \\ \hline

$^{50}$Ti+$^{248}$Cm    &   -185.52     &   229.00  &   12.25   &   3.75  & 3.9 & 219.88 & 216.39 &  34.36  &   4.41    \\
$^{50}$Ti+$^{247}$Bk    &   -191.42     &   231.85  &   12.25   &   3.80  & 4.1 & 222.63 & 219.00 &  31.21  &   4.14    \\
$^{50}$Ti+$^{249}$Bk    &   -189.78     &   231.45  &   12.25   &   3.67  & 3.3 & 222.16 & 218.71 &  32.38  &   4.18    \\
$^{50}$Ti+$^{250}$Cf    &   -194.40     &   233.79  &   12.25   &   3.64  & 4.9 & 224.56 & 220.84 &  30.16  &   3.98    \\
$^{50}$Ti+$^{252}$Cf    &   -193.02     &   233.23  &   12.25   &   3.53  & 4.2 & 223.97 & 220.33 &  30.95  &   3.96    \\
$^{50}$Ti+$^{254}$Cf    &   -191.92     &   232.67  &   12.5    &   4.38  & 3.6 & 223.38 & 219.81 &  31.46  &   3.97    \\
$^{50}$Ti+$^{252}$Es    &   -197.90     &   235.72  &   12.25   &   3.52  & 4.3 & 226.37 & 222.67 &  28.47  &   3.71    \\
$^{50}$Ti+$^{254}$Es    &   -196.81     &   235.24  &   12.5    &   4.39  & 3.8 & 225.86 & 222.21 &  29.05  &   3.69    \\

\hline\hline

\end{tabular}
\end{table}

In addition to the reactions induced by $^{48}$Ca leading to
superheavy nuclei, reactions with $^{36}$S, $^{37}$Cl, $^{48}$Ca
and $^{50}$Ti bombarding on $^{248}$Cm, $^{247,249}$Bk,
$^{250,252,254}$Cf and $^{252,254}$Es are also studied and all
relevant parameters for the entrance-channel capture barriers for
those fusion reactions are listed in Table 2. The table gives the
Q-value for the reactions, the barrier height $B_0$, position
$R_0$ of the barrier, curvature at the top of the barrier
expressed by $\hbar \omega_0$, factor $\gamma$ of structure
effects, the mean value $B_{\rm mean}$ of the barrier, the most
probable barrier energy $B_{\rm m.p.}$, the excitation energy of
compound nucleus $E^{*}_{CN}$ when $E_{\rm c.m.}=B_{\rm mean}$,
and the depth of capture pocket $B_0-B_s$ (or called quasi-fission
barrier height\cite{Ada97}, here $B_s$ denotes the value at the
bottom of the pocket, see Fig.1). Comparing the data from
different reactions one can find that the reactions with $^{37}$Cl
induce relatively higher excitation energies $E^{*}_{CN}$ and
those with $^{48}$Ca and $^{50}$Ti produce relatively lower
excitation energies when $E_{\rm c.m.}=B_{\rm mean}$. So $^{48}$Ca
and $^{50}$Ti induced reactions can be considered as good
candidates of cold fusion reaction for producing superheavy nuclei
from the point of a low excitation energy of the compound nuclei.
Here we have not studied the orientation effect of deformed
target, which has significant effect on fusion barrier height and
the compactness of the fusion reactions. Recently, compactness of
the $^{48}$Ca induced hot fusion reactions was studied in which it
was shown that $^{48}$Ca induced reactions on various actinides
were the best cold fusion reactions with optimum orientations of
the hot fusion process\cite{Gup06}. By comparing the depths of the
capture pockets for different reactions we find that the depth
decreases with increase of the proton number of the projectile
nuclei. We know that the shallower the pocket is, the stronger the
quasi-fission is. So the projectile $^{36}$S inducing capture
reactions is  more favorable for the small quasi-fission
probabilities of those reactions. By using this table we can
easily calculate the capture cross sections by eqs.$(15)-(17)$ for
all reactions listed. Fig.10 shows the calculated capture
excitation functions for the systems $^{36}$S+$^{254}$Es and
$^{48}$Ca+$^{254}$Es with our approach by using the data from
Table 2. In addition, the entrance-channel capture barriers of the
reactions $^{64}$Ni+$^{238}$U, $^{58}$Fe+$^{244}$Pu,
$^{54}$Cr+$^{248}$Cm and $^{50}$Ti+$^{252}$Cf which lead to the
same compound nucleus with $Z=120$ and $N=182$ are calculated and
listed in Table 3. Table 2 and Table 3 provide us with very useful
information for choosing an optimal combination of projectile and
target and suitable incident beam energies for producing
superheavy nuclei for unmeasured massive fusion reactions.

\begin{table}
\caption{ The same as Table 2, but for reactions
$^{64}$Ni+$^{238}$U, $^{58}$Fe+$^{244}$Pu, $^{54}$Cr+$^{248}$Cm
and $^{50}$Ti+$^{252}$Cf.}
\begin{tabular}{cccccccccc}
\hline \hline
  Reaction & Q$(MeV)$ & $ B_0(MeV)$ & $R_0(fm)$ & $\hbar \omega_0(MeV)$ & $\gamma$ & $B_{\rm mean}(MeV)$  & $B_{\rm m.p.}(MeV)$ & $E^{\ast}_{CN}(MeV)$ & $B_0$-$B_s$  \\ \hline

$^{64}$Ni+$^{238}$U     &   -237.41     &   276.01  &   12.5    &   4.61  & 7.1 & 265.26 & 260.73 &  27.85  &   1.79    \\
$^{58}$Fe+$^{244}$Pu    &   -219.97     &   262.88  &   12.25   &   3.93  & 1.0 & 251.56 & 248.80 &  31.60  &   2.56    \\
$^{54}$Cr+$^{248}$Cm    &   -207.16     &   248.52  &   12.25   &   4.20  & 3.0 & 238.51 & 234.86 &  31.35  &   3.26    \\
$^{50}$Ti+$^{252}$Cf    &   -193.02     &   233.23  &   12.25   &   3.53  & 4.2 & 223.97 & 220.33 &  30.95  &   3.96    \\

\hline\hline

\end{tabular}
\end{table}

\begin{center}
\textbf{V. CONCLUSION AND DISCUSSION}
\end{center}

In this work, the Skyrme energy-density functional approach has
been applied to study massive heavy-ion fusion reactions,
especially those leading to superheavy nuclei. Based on the
barriers calculated with the Skyrme energy-density functional, we
propose the parameterized barrier distributions to effectively
taken into account the multi-dimensional character of the
realistic barrier. A large number of heavy-ion fusion reactions
have been studied systematically. The average deviations of fusion
cross sections at energies near and above the barriers from
experimental data are less than 0.05 for $92\%$ of 76 fusion
reactions with $Z_1Z_2<1200$. Massive fusion reactions, for
example, the $^{238}$U-induced reactions and the
$^{48}$Ca+$^{208}$Pb have been studied and their capture
excitation functions have been reproduced well. The influence of
the structure effects in the reaction partners on the capture
cross sections are studied by using parameter $\gamma$ in our
model. To search the most favorable condition for the synthesis of
superheavy nuclei, the optimal balance between the capture cross
section and the excitation energy of the formed compound nuclei is
studied by comparing the fusion reactions induced by the
double-magic nucleus $^{48}$Ca and by $^{32}$S and $^{35}$Cl.
Based on this study, the 'threshold-like' behavior of excitation
function of capture cross sections with respect to incident beam
energy has been explored and possible values of this 'threshold'
for reactions mainly induced by $^{48}$Ca are given. Finally, we
have further studied the capture reactions leading to superheavy
nuclei such as $^{36}$S, $^{37}$Cl, $^{48}$Ca and $^{50}$Ti
bombarding on $^{248}$Cm, $^{247,249}$Bk, $^{250,252,254}$Cf and
$^{252,254}$Es, and as well as the reactions $^{64}$Ni+$^{238}$U,
$^{58}$Fe+$^{244}$Pu, $^{54}$Cr+$^{248}$Cm and
$^{50}$Ti+$^{252}$Cf which lead to the same compound nucleus with
$Z=120$ and $N=182$. The relevant parameters for calculating the
capture cross section of these reactions have been provided which
are helpful for the study of unmeasured massive fusion reactions.
Especially, we predicted optimal fusion configuration and suitable
incident beam energies for the synthesis of superheavy nuclei.

We notice that the deformation and orientation of colliding nuclei
have a very significant role on fusion reactions. In
\cite{Iwa96,Gup05} the effect of deformation and orientation on
the barrier hight and the compactness of fusion reactions were
investigated systematically. However, this kind study is beyond
the scope of present work. We have only made preliminary
calculations of the potential barrier for $^{48}$Ca+$^{248}$Cm
with the deformation and orientation of $^{248}$Cm taken into
account in the entrance channel. For this reaction the lowest
barrier is obtained for the orientation $\Theta=0^{\, \circ}$,
i.e. when $^{48}$Ca touches the tip of deformed $^{248}$Cm target,
while the highest barrier is obtained for $\Theta=90^{\, \circ}$,
when $^{48}$Ca touches the side. The lowest barrier obtained for
$\Theta=0^{\, \circ}$ is a little bit lower than the most probable
barrier height $B_{\rm m.p.}$ of this reaction given in Table 1
and the barrier distribution due to the orientation of $^{248}$Cm
is close to the effective weighting function $D_{\rm eff} (B)$
which is for describing the capture process of the reaction if
assuming the orientation probability decreases gradually from
$0^{\, \circ}$ to $90^{\, \circ}$. So the deformation effects seem
to be partly involved in the parameterized barrier distribution
functions. The study on this aspect is in progress.

\section{Acknowledgements}

This work is supported by Alexander von Humboldt Foundation and
National Natural Science Foundation of China, No. 10235030,
10235020. Wang is grateful to Prof. Enguang Zhao and Prof. Junqing
Li for fruitful discussions.

\end{document}